# Design of A Low-Latency and Parallelizable SVD Dataflow Architecture on FPGA

Fangqiang Du, Sixuan Chong, Zixuan Huang, Rui Qin, Fengnan Mi, Caibao Hu, Jiangang Chen

*Abstract*—Singular value decomposition (SVD) is widely used for dimensionality reduction and noise suppression, and it plays a pivotal role in numerous scientific and engineering applications. As the dimensions of the matrix grow rapidly, the computational cost increases significantly, posing a serious challenge to the efficiency of data analysis and signal processing systems, especially in time-sensitive scenarios involving large-scale datasets. Although various dedicated hardware architectures have been proposed to accelerate the computation of intensive SVD, many of these designs suffer from limited scalability and high consumption of on-chip memory resources. Moreover, they typically overlook the computational and data transfer challenges associated with SVD, making them unsuitable for real-time processing of large-scale data stream matrices in embedded systems. In this paper, we propose a Data Stream-Based SVD processing algorithm (DSB Jacobi), which significantly reduces on-chip BRAM usage while improving computational speed, offering a practical solution for real-time SVD computation of large-scale data streams. Compared to previous works, our experimental results indicate that the proposed method reduces on-chip RAM consumption by 41.5% and improves computational efficiency by a factor of 23.

## I. INTRODUCTION

SINGULAR value decomposition (SVD) is a fundamental operation in linear algebra and, as a powerful mathematical tool, has become a key theoretical foundation for various emerging applications in embedded systems. By decomposing a matrix into three specifically structured submatrices, SVD effectively captures the intrinsic structure and latent features of the data, establishing it as one of the core algorithms in modern signal processing. It demonstrates significant advantages in tasks such as principal component analysis[1], dimensionality reduction[2], and noise suppression[3], and is widely applied in practical scenarios including communication systems[4], [5], [6], image compression[7], [8], [9], and ultrasound image filtering[10], [11], [12].

Despite the significant theoretical value and practical potential of SVD in scientific computing and engineering applications, its inherent computational and implementation limitations have constrained its deployment in real-world systems. First, SVD involves high computational complexity[13], [14]. For a matrix satisfying $m \geq n$, the time complexity is $O(mn^2)$, leading to substantial computational overhead in large-scale data processing scenarios and making real-time processing difficult to achieve. Secondly, SVD has very high storage requirements[14], [15]. The full decomposition requires storing three dense matrices, which can consume a large amount of memory for large-scale data and limit its use in resource-constrained environments, such as embedded systems. In addition, because SVD involves intricate computation steps and offers limited inherent parallelism[13], [14], achieving efficient parallel realization remains challenging, especially during acceleration or hardware implementation. These issues collectively present significant challenges for designing SVD processors in time-sensitive applications.

In recent years, substantial research efforts have been devoted to the parallelization and real-time processing of SVD, leading to the development of numerous hardware-oriented SVD computation methods. Reference[13] proposed a general FPGA-based hardware architecture that performs SVD computation for large-scale $m \times n$ matrices using the Hestenes method combined with one-sided Jacobi rotations. Reference[15] introduced the Maximum Data Sharing ordering, which effectively reduces costly off-chip data transfers and bandwidth requirements by maximizing on-chip data reuse. Reference[16] proposed a parallel one-sided Jacobi SVD method with an adjustable block size, introducing a column-block-based SVD computation strategy. Reference[14] presented a BCV Jacobi algorithm for efficiently computing the SVD of matrices of arbitrary size. Previous studies have primarily focused on parallelizing the SVD algorithm. However, most existing approaches exhibit limited scalability and struggle to achieve efficient performance on resource-constrained FPGA platforms and in real-time processing scenarios.

To solve the problem mentioned above, this study primarily concentrates on the following aspects:

(1) We introduce a novel Data Stream-Based SVD processing algorithm (DSB Jacobi), which transforms the column-pair-based orthogonalization in the traditional Hestenes method into a row-pair-based approach, better aligning with

This work was partially supported by the Science and Technology Commission of Shanghai (Grant No. 22DZ2229004, 22JC1403603, 21Y11902500); the Key Research & Development Project of Zhejiang Province (2024C03240); Jilin Province science and technology development plan project (Grant No. 20230204094YY); 2022 "Chunhui Plan" cooperative scientific research project of the Ministry of Education. *(Corresponding author: Jiangang Chen, Caibao Hu).*

Fangqiang Du is with the East China Normal University, Shanghai 200241, China (e-mail: 71265904012@stu.ecnu.edu.cn).
Jiangang Chen is with the East China Normal University, Shanghai 200241, China (e-mail: jgchen@cee.ecnu.edu.cn).
Caibao Hu is with the Department of Critical Care Medicine, Zhejiang Hospital, No. 12, Lingyin Road, Xihu District, Hangzhou, Zhejiang 310013, China (e-mail: zjicu1996@163.com).
Sixuan Chong, Zixuan Huang, Rui Qin are with the East China Normal University, Shanghai 200241, China (e-mail: 740520291@qq.com; 13246009616@163.com; 51265904015@stu.ecnu.edu.cn).
Fengnan Mi is with the Shanghai Publishing and Printing College, Shanghai 200093, China (e-mail: mifengnan@aliyun.com).

the dataflow characteristics of streaming processing. This method significantly reduces the storage requirements for intermediate data and effectively shortens the SVD computation time, offering a new possibility for real-time SVD.

(2) We propose a RAM resource sharing strategy, in which all data buffering operations within a single Processing Unit (PU) share the same RAM resource. This approach significantly reduces on-chip memory usage during SVD computation, making it feasible to implement the DSB Jacobi algorithm on a resource-constrained FPGA device. In addition, by leveraging the parallelism of the FPGA and introducing a pipelined data flow, this architecture compensates for the latency introduced by shared memory access, achieving an overall throughput improvement.

(3) We present a flexible hardware framework in which the core algorithms are packaged into modular functional blocks. Such an arrangement allows the architecture to be readily tailored to different PU configurations with only minor code adjustments, thereby simplifying the process of algorithm migration and reducing development overhead. Moreover, the architecture offers excellent scalability, allowing for the flexible configuration of the number of PU modules based on the resource availability of different FPGA devices, thereby providing an efficient and practical solution for a wide range of application scenarios.

## II. PROPOSED ARCHITECTURE

### A. SVD

The SVD algorithm is used to decompose an arbitrary matrix $A_{m\times n}$ (without loss of generality, $m \geq n$) into the product of three submatrices $(U, \Sigma, V)$. The decomposition can be expressed as follows[17]:

$$A_{m\times n} = U_{m\times n} \times \Sigma_{n\times n} \times V_{n\times n}^T \quad (1)$$

Here, $U_{m\times n}$ corresponds to the left singular matrix, while $V_{n\times n}$ corresponds to the right singular matrix, respectively, and both are unitary, satisfying $UU^T = I_m$ and $VV^T = I_n$. The matrix $\Sigma_{n\times n}$ is diagonal, with diagonal elements denoting the singular values of the original matrix $A$, i.e., $\Sigma = \text{diag}(\sigma_1, \sigma_2, \ldots, \sigma_n)$.

### B. Hestenes Method

The Hestenes method, also known as the implicit one-sided Jacobi SVD[18], is based on performing a series of Jacobi rotations on the matrix $A_{m\times n}$. The fundamental principle underlying the one-sided Jacobi algorithm can be expressed as follows:

$$B = AV = A(J_1 J_2 J_3 \ldots) \quad (2)$$

Matrix $B$ with pairwise orthogonal column vectors is obtained, such that $b_i^T b_j = 0$. The matrix $B$ is then normalized to yield [20]:

$$B = U \times \Sigma \quad (3)$$

$\Sigma$ is a diagonal matrix, expressed as $\Sigma = \text{diag}(\sigma_1, \sigma_2 \ldots \sigma_{n-1}, \sigma_n)$, where $\sigma_i = b_i^T b_i$. Since the matrix $V = J_1 J_2 J_3 \ldots$ is formed by the product of a sequence of Jacobi rotation, it follows that the matrix $V$ is orthogonal. By rearranging (2) and (3), the SVD algorithm of matrix $A$ can be expressed as follows:

$$A = U\Sigma V^T \quad (4)$$

### C. DSB-Jacobi Algorithm

Building upon the traditional Hestenes-Jacobi method, this paper proposes a DSB Jacobi algorithm. The matrix $A_{m\times n}$ is evenly partitioned by rows into series blocks, each containing $NumOfPu$ rows. Without loss of generality, this paper assumes that the matrix is partitioned into several non-overlapping submodules that collectively cover the entire matrix. Based on this assumption, the detailed execution procedure of the DSB Jacobi is presented in Algorithm 1.

---
**Algorithm 1: DSB Jacobi**

**Input:** $A_{m\times n}$, $NumOfConv$, $NumOfPu$
**Output:** $U_{m\times n}$, $\Sigma_{n\times n}$, $V_{n\times n}$

1  $U = A^T$
2  $[m, n] = \text{size}(A)$
3  $conv\_count = NumOfConv$
4  $NumOfSweep = \frac{n}{NumOfPu}$
5  $V = eye(n)$
6  **while** $(conv\_count > 0)$ **do**
7     **for** $k = 1$ to $(NumOfSweep - 1)$ **do**
8        **for** $i = ((k-1) \times NumOfPu + 1)$ to $(k \times NumOfPu - 1)$ **do**
9           **for** $j = (i+1)$ to $(k \times NumOfPu)$ **do**
10              Calculate: $\alpha, \beta, \gamma$
11              Calculate: $\sin\theta, \cos\theta$
12              Update Matrix of $U, V$
13          **end**
14        **end**
15     **end**
16     $conv\_count = conv\_count - 1$
17  **end**
   /* Update Matrix of $U, \Sigma$ */
18  **for** $j = 1$ to $n$ **do**
19     $\sigma_j = norm(U_j)$
20     $U_{,j} = \frac{U_{,j}}{\sigma_j}$
21  **end**
   /* Generate Matrix of $U, \Sigma, V$ */
22  $U = U^T$
23  $\Sigma = diag(\sigma_j)$
24  $V = V^T$

---

## III. SYSTEM DESIGN

### A. System Architecture

The system architecture, illustrated in Fig. 1, mainly consists of two main components: the TestBench and the svd_kernel. The TestBench module serves as the system simulation module, while the svd_kernel module is a synthesizable RTL module responsible for performing SVD decomposition on the FPGA. Previous work generated only the S matrix [15], while this algorithm has been improved to output the U, S, and V matrices simultaneously.

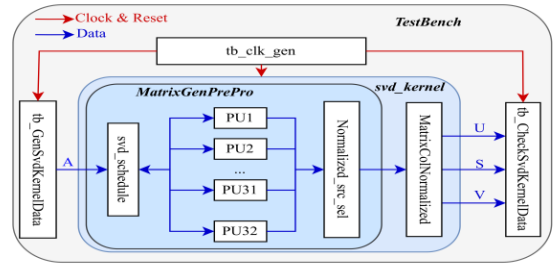

**Fig. 1.** The system architecture of SVD based on an FPGA.

### B. Cyclic Scheduling

The data scheduling algorithm primarily handles data management for each Jacobi operation across multiple PU modules. Taking an example of eight rows and four PU modules, the scheduling process is illustrated in Fig. 2. The data processing flow was divided into four stages. The first stage corresponds to row data buffering (Fig. 2(a)), while the second through fourth stages represent data scheduling and updating (Fig. 2 (b–d)). In each stage, the numbers 1,2,3… denote the row indices of the processed data.

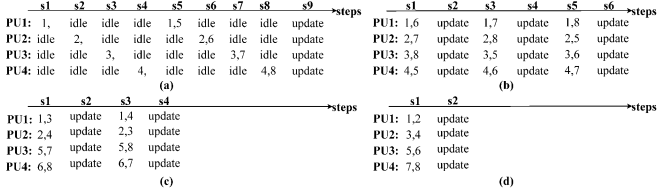

**Fig. 2.** Cyclic data scheduling process on four PU modules: (a) stage one, (b) stage two, (c) stage three, (d) stage four.

Compared to the data scheduling algorithm in existing studies[14], this work introduces two key improvements. First, we modified the traditional SVD computation from column-wise grouping to row-wise grouping, which significantly improves data access efficiency. This row-based processing approach aligns better with common dataflow architectures, making it particularly suitable for applications where data is input row by row, such as image processing. Second, the data scheduling mechanism between PUs has been simplified, enabling tighter timing control and higher overall scheduling efficiency.

*C. Processing Unit*

Fig. 3 illustrates the architecture of the PU module, comprising three submodules: pu_ram_ctrl, param_gen, and update_matrix. The pu_ram_ctrl submodule is responsible for buffering and scheduling control of row data, param_gen generates the sine and cosine parameters, and update_matrix performs the updates of the $U$ and $V$ matrices.

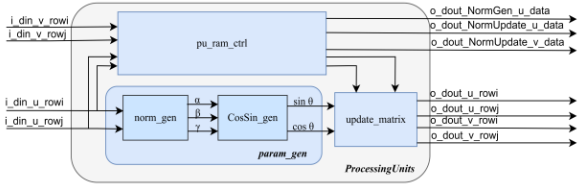

**Fig. 3.** The architecture of the PU module.

The PU module performs the following operations. First, two lines of input data are simultaneously sent to the param_gen submodule, which calculates the corresponding $sin\theta$ and $cos\theta$ parameters. Meanwhile, these two lines of data are buffered in the pu_ram_ctrl submodule. The computed $sin\theta$ and $cos\theta$ values, together with the buffered data rows, are then passed to the update_matrix submodule, which executes the Jacobi rotation and outputs the updated rows.

To reduce RAM usage, the design implements the following optimization:

(1) Each PU module instantiates only four RAM blocks to store the $i$-th and $j$-th rows of the $U$ and $V$ matrices, respectively. Taking the $U$ matrix as an example, two of these RAMs are reused across multiple computation stages. Initially, they buffer the input row data to generate the $sin\theta$ and $cos\theta$ parameters as well as to update the matrix elements. Subsequently, they buffer the results from the preceding PU computation to serve as input for the current iteration. This row-level reuse strategy effectively reduces the on-chip memory demand, resulting in nearly a one-third decrease in overall RAM utilization.

(2) After all PU modules finish updating the rows of the $U$ and $V$ matrices, the system performs two pipelined memory reads from the buffered data to obtain the final U, S, and V components. During the first read, the norm is computed to form the $S$ matrix, while the second read assembles the $U$ matrix. This pipeline processing strategy enables parallel computation and also mitigates the issue of increased time delays caused by shared RAM, thereby enhancing overall computational efficiency.

## IV. EXPERIMENTS AND RESULTS

To assess the performance of the DSB Jacobi algorithm, we first compare it with the conventional Hestenes Jacobi algorithm using MATLAB. The consistency of the results validates the correctness of the DSB Jacobi algorithm. Subsequently, we evaluate the computational performance of the DSB Jacobi algorithm with varying numbers of iterations and PU modules. Finally, the algorithm is implemented at the RTL level on the Xilinx XCKU060-FFVA1517 FPGA platform.

*A. Performance Analysis*

Accuracy and runtime are two primary metrics for assessing the performance of SVD computation. In this work, the matrix norm is used to quantify the decomposition error and the orthogonality errors of the U and V matrices.
The SVD computation error is defined as follows:
$$E_{svd} = A - U\Sigma V^T \tag{5}$$
The norm-based definition of the SVD computation error is given as follows:
$$NORM_{error\_svd} = \sum_{i=1}^{m}\sum_{j=1}^{n}(E_{svd})_{i,j}^2 \tag{6}$$
The orthogonality errors of the matrices $U$ and $V$ are defined as follows:
$$E_{uq} = UU^T - I_m \tag{7}$$
$$E_{vq} = VV^T - I_n \tag{8}$$
The norm-based definition of the orthogonality errors for the matrices U and V is given as follows:
$$NORM_{error\_uq} = \sum_{i=1}^{m}\sum_{j=1}^{m}(E_{uq})_{i,j}^2 \tag{9}$$
$$NORM_{error\_vq} = \sum_{i=1}^{n}\sum_{j=1}^{n}(E_{vq})_{i,j}^2 \tag{10}$$

*B. Comparisons with Hestenes-Jacobi*

The DSB Jacobi and the conventional Hestenes Jacobi were implemented in MATLAB. Their computational errors were comparatively analyzed, as shown in Fig. 4. Fig. 4(a), (b), and (c) illustrate the comparison of the SVD decomposition error norm, the orthogonality error norm of the $U$ matrix, and the $V$ matrix, respectively, under a single iteration. According to the error metrics, the two algorithms yield identical results, indicating a strong agreement in their computational outputs. This result validates the numerical accuracy of the proposed DSB Jacobi algorithm and the feasibility of the architecture.

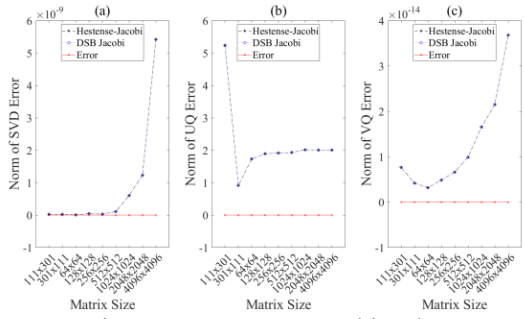

**Fig. 4.** Comparison Between DSB Jacobi and Hestenes Method Results: (a) the SVD decomposition error norm, (b) the orthogonality error norm of the U matrix, (c) the orthogonality error norm of the V matrix.

*C. Analysis of DSB Jacobi Algorithm Results with Varying Number of Iterations*

To study the impact of the number of iterations on performance, a full DSB Jacobi was performed on a $256 \times 256$ matrix in MATLAB. Fig. 5 shows the computation time, the norm of the SVD reconstruction error, and the orthogonality error norms of the U and V matrices. The reconstruction error remains below $10^{-9}$ across all iterations (Fig. 5(b)), while the orthogonality error of the V matrix stays below $10^{-14}$ (Fig. 5(d)), confirming the high numerical precision of the algorithm. As the number of iterations increases, computation time grows linearly (Fig. 5(a)), and the U-matrix orthogonality improves, dropping below $10^{-4}$ beyond ten iterations (Fig. 5(c)). Thus, the iteration counts can be flexibly selected to balance computational efficiency and orthogonality precision in practical applications.

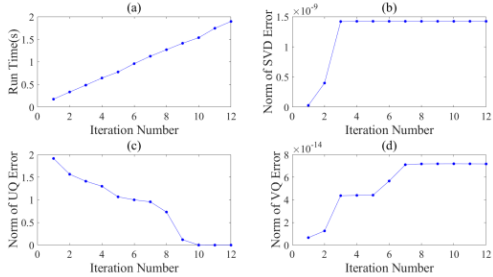

**Fig. 5.** Results with Varying Number of Iterations: (a) the computation time, (b) the SVD decomposition error norm, (c) the orthogonality error norm of the U matrix, (d) the orthogonality error norm of the V matrix.

*D. Analysis of DSB Jacobi Algorithm Results with Varying Row Number of PU*

To evaluate the impact of varying row number of PU module on computational results, DSB Jacobi architectures with different PU configurations were implemented in MATLAB and evaluated across various matrix sizes. The metrics include computation time, SVD decomposition error norm, and the orthogonality error norms of the U and V matrices. As shown in Fig. 6, all configurations maintain SVD decomposition error norms below $10^{-10}$ (Fig. 6(b)) and V-matrix orthogonality errors below $10^{-14}$ (Fig. 6(d)), demonstrating high numerical accuracy and strong orthogonality. Increasing the number of rows per PU slightly reduces the U-matrix orthogonality error (Fig. 6(c)) but increases computation time (Fig. 6(a)). Therefore, the PU configuration can be flexibly chosen according to available resources and latency constraints to balance accuracy and efficiency.

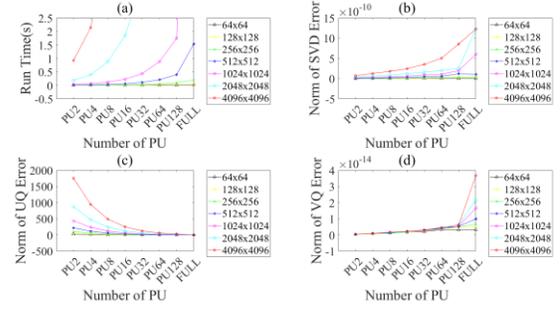

**Fig. 6.** Results with Varying Row number of PU: (a) the computation time, (b) the SVD decomposition error norm, (c) the orthogonality error norm of the U matrix, (d) the orthogonality error norm of the V matrix.

*E. Analysis of Execution Times and Resource Utilization on FPGA*

The execution time and resource utilization of the proposed DSB Jacobi on FPGA are summarized in Tables I and II, respectively. As shown in Table I, the SVD computation time remains nearly constant across different PU architectures for matrices of the same size, indicating that the PU structure is not the dominant factor influencing computational latency. Hence, in real-time applications, any PU configuration can be flexibly selected according to system requirements without compromising execution efficiency. Table II shows that FPGA resource utilization increases proportionally with the number of rows in each PU module. Therefore, it is necessary to select an appropriate PU architecture based on the available FPGA resources to maintain an effective balance between computational performance and hardware efficiency.

TABLE I
EXECUTION TIMES (MILLISECOND) OF DSB JACOBI

| Matrix | PU2 | PU4 | PU8 | PU16 | PU32 |
|---|---|---|---|---|---|
| $128 \times 128$ | 0.499 | 0.513 | 0.521 | 0.520 | 0.537 |
| $256 \times 256$ | 1.531 | 1.537 | 1.544 | 1.532 | 1.565 |
| $512 \times 512$ | 5.192 | 5.123 | 5.096 | 5.030 | 5.095 |
| $1024 \times 1024$ | 18.903 | 18.437 | 18.220 | 17.924 | 18.055 |
| $2048 \times 2048$ | 71.885 | 69.642 | 68.552 | 64.425 | 67.568 |
| $4096 \times 4096$ | 280.084 | 270.356 | 265.558 | 260.440 | 260.964 |

TABLE II
RESOURCE UTILIZATION OF DSB JACOBI

| Resource | PU2 | PU4 | PU8 | PU16 | PU32 |
|---|---|---|---|---|---|
| LUT | 11K | 22K | 43K | 85.6K | 170.4K |
| FF | 21.5K | 39.5k | 75.5K | 147.4K | 291.2K |
| BRAM | 19 | 38 | 76 | 152 | 304 |
| DSP | 133 | 265 | 529 | 1057 | 2113 |

*F. Comparisons with Prior Studies*

For a single iteration, the performance comparison between the proposed DSB Jacobi algorithm and previous studies is presented in Table III. The results indicate that the proposed method significantly improves computational efficiency while reducing hardware resource usage. For example, for a $4096 \times 4096$ matrix, the execution time reported in[15] is 12.2464 seconds, which is insufficient for real-time applications. Similarly, the design in[14] requires 6.0259 seconds, whereas our implementation completes the same task in only 261 milliseconds, achieving approximately a 23-fold speedup. Additionally, BRAM utilization decreases from 519.5 in[14] to 304 in our design, representing a 41.5% reduction. These results confirm that the proposed method significantly improves computational throughput and resource usage on FPGA platforms.

TABLE III
EXECUTION TIMES (SECONDS) AND RESOURCE UTILIZATION WITH EXISTING STUDIES

|  | [15] | [14] | This Work |
|---|---|---|---|
| $128 \times 128$ | 0.0014 | 0.0002 | 0.0005 |
| $256 \times 256$ | 0.0066 | 0.0019 | 0.0016 |
| $512 \times 512$ | 0.0347 | 0.0138 | 0.0051 |
| $1024 \times 1024$ | 0.2285 | 0.1020 | 0.0181 |
| $2048 \times 2048$ | 1.6299 | 0.7752 | 0.0676 |
| $4096 \times 4096$ | 12.2464 | 6.0259 | 0.2610 |
| Platform | ZC706 | XC7V690T | XCKU060 |
| Clock | 150Mhz | 200Mhz | 200Mhz |
| LUT | 92K | 212K | 170.4K |
| DSP | 712 | 1602 | 2113 |
| BRAM | 284 | 519.5 | 304 |

## V. DISCUSSION

This study demonstrates that parallel and efficient SVD computation can be effectively realized even on resource-constrained FPGA devices. The results show that increasing the number of rows in the PU module in MATLAB leads to longer computation times (Fig. 6(a)). In contrast, the FPGA, due to its parallel advantage, shows that the SVD computation time is independent of the PU architecture (Table I), confirming the efficiency of the proposed design. As shown in Table II, using fewer rows per PU reduces FPGA resource utilization, but it also slightly decreases computation accuracy (Fig. 6(c)). Conversely, increasing the iteration count improves accuracy (Fig. 5(c)) at the cost of longer execution time (Fig. 5(a)). Therefore, high-precision SVD decomposition can be achieved on resource-limited devices by selecting a smaller PU module while increasing the number of iterations appropriately. Compared to previous approaches that require substantial hardware resources and longer runtimes, the proposed design enables practical deployment on a compact FPGA device, supporting real-time large-scale matrix processing. Although the current evaluation is based on simulation data, future work will extend this architecture to real-time ultrasound image filtering, further validating its applicability.

## VI. CONCLUSION

This paper presents a fully hardware-based SVD solver implementing the DSB Jacobi algorithm. Compared with prior designs, the proposed architecture offers notable advantages in three aspects. (a) Efficiency: Orthogonalization is performed on row pairs, which reduces storage and data transfer overhead, thereby improving real-time performance. (b) Structural Simplicity: The iterative scheduling algorithm features a simple structure and precise data flow, which facilitates FPGA timing convergence design and thereby enhances implementation reliability. (c) Flexibility: This architecture is highly scalable and can configure different PU architectures based on the available FPGA resources, providing a flexible and resource-efficient solution for various application scenarios.